\documentclass[10pt,letterpaper,twocolumn]{article} 

\usepackage{ol2}
\usepackage[draft,implicit=false]{hyperref}
\usepackage{amsmath}
\usepackage{upgreek}
\begin{document}

\twocolumn[ 

\title{Active Faraday optical frequency standards}


\author{Wei Zhuang, and Jingbiao Chen$^{*}$}

\address{
State Key Laboratory of Advanced Optical Communication Systems and Networks,  Institute of Quantum Electronics, \\School of Electronics Engineering $\&$ Computer Science, Peking University, Beijing 100871, China\\
$^*$Corresponding author: jbchen@pku.edu.cn
}

\begin{abstract}We propose the mechanism of active Faraday optical clock, and experimentally demonstrate active Faraday optical frequency standards based on 852 nm narrow bandwidth Faraday atomic filter by the method of velocity-selective optical pumping of cesium vapor. The center frequency of the active Faraday optical frequency standards is determined by the cesium 6 $^{2}S_{1/2}$ $F$ = 4 to 6 $^{2}P_{3/2}$ $F'$ = 4 and 5 crossover transition line. The optical heterodyne beat between two similar independent setups shows that the frequency linewidth reaches 996(26) Hz, which is 5.3 $\times$ 10$^{3}$ times smaller than the natural linewidth of the cesium 852 nm transition line. The maximum emitted light power reaches 75 $\upmu$W. The active Faraday optical frequency standards reported here have advantages of narrow linewidth and reduced cavity pulling, which can readily be extended to other atomic transition lines of alkali and alkaline-earth metal atoms trapped in optical lattices at magic wavelengths, making it useful for new generation of optical atomic clocks.\end{abstract}


] 

\noindent The optical frequency standards based on neutral atoms in optical lattices~\cite{Takamoto05, Hinkley13, Ye14, Katori} have shown better stability and accuracy at 10$^{-18}$ level. However, the performance of the optical lattice clocks is still limited by the Brownian thermo-mechanical noise of high-finesse optical cavities for frequency stabilization of clock lasers~\cite{Jiang11, Kessler12}. Using atoms with narrow linewidth clock transitions and a bad-cavity to make an active optical clock \cite{Chen05, Chen09, Chen08, Chen11, Kazakov13, Meiser09, Meiser10, Bohnet12}, can greatly reduce the influence of the  mechanical or thermal vibrations of the cavity mirrors on the emitted optical frequency. A spectral linewidth of just 1 mHz and a potential stability of two orders of magnitude better could be expected \cite{Chen05, Chen09, Chen08, Chen11, Kazakov13, Meiser09, Meiser10, Bohnet12, Sterr09}.

The active optical clocks \cite{Chen05, Chen09, Chen08} work in the bad-cavity regime, which corresponds to the condition that the cavity mode bandwidth $\Gamma_{c} $ is larger than the gain bandwidth $\gamma_{a} $. The bad-cavity laser based on HeNe \cite{Kuppens94} or HeXe \cite{Kuppens96} infrared gas lasers with  $\Gamma_{c}/\gamma_{a}$  equals 1.4 was reported. The laser using rubidium two-photon Raman transition realized $\Gamma_{c}/\gamma_{a}$ of 5$\times$10$^{4}$ \cite{Bohnet12} and the Gaussian full-width at half-maximum (FWHM) was measured to be 350(25) Hz relative to the Raman dressing laser.

In this Letter, we introduce the concept of the active Faraday optical clock, in which the optical emission frequency is determined by the center frequency of the Faraday atomic filter \cite{Faraday, Turner02, Cer09, Popescu10, Liu12, Wang12} when working in bad-cavity regime. The optical gain can be provided by Ti:Sapphire, dye or semiconductor materials. We experimentally demonstrate an active Faraday frequency standard using Cs 852 nm narrow bandwidth Faraday atomic filter in an extended bad-cavity of a laser diode. The maximum emitted light power reaches 75 $\upmu$W and the frequency is determined by Cs 6 $^{2}S_{1/2}$ $F$ = 4 to 6 $^{2}P_{3/2}$ $F'$ = 4 and 5 crossover transition line with FWHM measured to be 996(26) Hz. The prospective active Faraday optical frequency standard would also employ Faraday alkali-earth atomic filter with ultra-narrow bandwidth, like strontium 689 nm and calcium 657 nm clock transitions in optical-lattice based Faraday atomic filter, with weak optical feedback from optical resonant cavity.

A schematic of the experimental setup is shown in Fig. 1. The active Faraday optical frequency standard shown in the black box consists of a laser diode (LD$_{1}$) and a Faraday atomic filter in the extended cavity between LD$_{1}$ and mirror M$_{0}$. The output fluorescence of laser diode (LD$_{1}$) with anti-reflective coatings (Eagleyard EYP-RWE-0860-06010-1500-SOT02-0000) is collimated by a aspheric lens L$_{1}$. The plane mirror M$_{0}$ with a reflectivity of 80\% used for optical feedback and output is about 35 cm away from the laser diode. A piezoelectric ceramic tube (PZT) is attached to the mirror M$_{0}$ to adjust the length of the extended cavity. The output of another similar laser diode (LD$_{2}$) is collimated by lens L$_{1}$, reflected by mirror M$_{1}$ and passes through the same Faraday atomic filter and feedback mirror M$_{0}$. These two output beams are combined by a beam splitter (BS$_{1}$, 50 : 50) after mirror M$_{0}$ for heterodyne detection. As the two output light frequencies are both determined by the Faraday atomic filter and equal to each other, one beam is shifted 113 MHz by an acousto-optic modulator (AOM)to facilitate analysis of beat signal. The overlapped beams are detected by an avalanche photodiode (APD$_{1}$) and analyzed using an rf spectrum analyzer. The other beams are further split by a polarization beam splitter (PBS$_{1}$) with one beam detected by a photodiode (PD$_{1}$) and another heterodyned with the pumping laser respectively.

\begin{figure}[htb]
\centerline{\includegraphics[width=8cm]{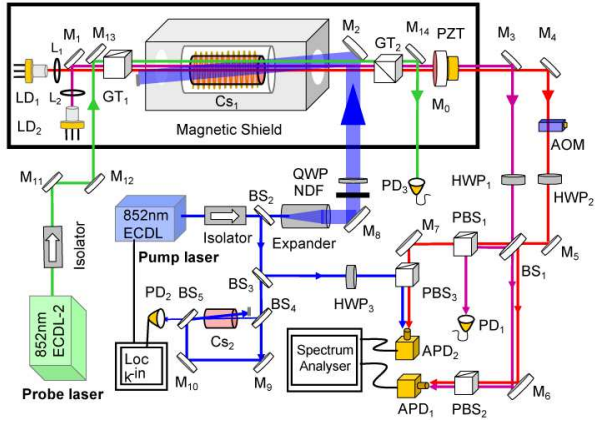}}
\caption{(Color online) The schematic of  the experimental setup. LD, laser diode; M, mirror; GT, Glan-Taylor prism; M$_{0}$, feedback extended cavity mirror; AOM, acousto-optic modulators; HWP, half-wave plate; QWP, quarter-wave plate; PBS, polarization beam splitter;  BS, beam splitter; PD, photodiode; APD, avalanche photodiode; ECDL, extended cavity diode laser; NDF, neutral density filter.}
\end{figure}

The narrow bandwidth Faraday atomic filter is realized by velocity-selective optical pumping of cesium vapor in a bias magnetic field \cite{Liu12, Wang12}. The Faraday atomic filter is composed of two polarization-orthogonal Glan-Taylor prisms (GT$_{1}$ and GT$_{2}$) with extinction ratio of 10$^{5}$ : 1, a cesium vapor cell, a pumping laser with circular polarization and a bias magnetic field. The pumping laser transfers velocity-selective Cs atoms to 6 $^{2}P_{3/2}$ $F'$ = 4 and 5 states. The bias magnetic field causes Zeeman splitting and magneto-optical rotation for linearly polarized probe beam \cite{Faraday} when interacting with velocity-selective atoms. Therefore, there is a transmission of probe beam passing through orthogonal prisms besides atomic medium with pumping laser.  In our experiment, a bias magnetic field about 10 Gs along the cell axis is applied to make Zeeman splitting match the frequency width of velocity-selective atoms to achieve high transmission of the Faraday atomic filter\cite{Wang12}.

The cesium vapor is contained in a cylindrical glass cell (Cs$_{1}$ in Fig.1) with a diameter of 25.4 mm and a length of 15 cm and temperature controlled at 30 $^{0}$C . A magnetic shielding chamber with two openings is used to reduce the influence of earth magnetic field. The pumping laser is provided by an extended cavity diode laser (ECDL) stabilized to Cs 6 $^{2}S_{1/2}$ $F$ = 4 to 6 $^{2}P_{3/2}$ $F'$ = 4 and 5 crossover transition line using saturation absorption spectrum (SAS) combined with electronic servo system. The pumping beam is expanded to overlap output light beam of LD$_{1}$ and LD$_{2}$, and circularly polarized before counter-propagating with them. The probe laser is utilized to measure the transmission of the Faraday atomic filter with an intensity of 0.5 mW/cm$^{2}$ lower than the saturation intensity 1.1 mW/cm$^{2}$. The laser frequency is swept over  2 GHz referenced to the Cs SAS and the transmitted light is recorded by a photodiode PD$_{3}$.

\begin{figure}[htb]
\centerline{\includegraphics[width=8cm]{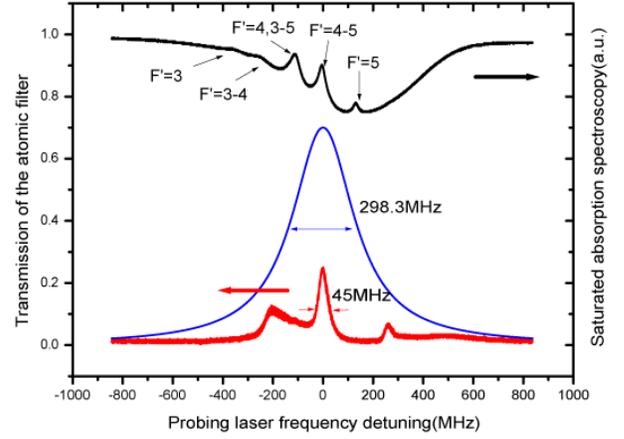}}
\caption{ (Color online) The transmittance spectrum of the Cs Faraday atomic filter at 6 $^{2}S_{1/2}$ $F$ = 4 to 6 $^{2}P_{3/2}$ $F'$ = 4 and 5 crossover transition line. }
\end{figure}

Figure 2 shows the transmission spectrum of the Faraday atomic filter. The center frequency of the filter coincides with the Cs 6 $^{2}S_{1/2}$ $F$ = 4 to 6 $^{2}P_{3/2}$ $F'$ = 4 and 5 crossover transition line. It possesses a transmission of 24.9(2)\% excluding the reflections of the vapor cell windows and the FWHM bandwidth is 45(2) MHz, which mainly results from pumping laser caused saturation broadening. The $Q$ factor of the whole extended cavity is calculated to be 1.18 $\times$ 10$^{6}$ and the FWHM of cavity mode is 298.3 MHz. In Fig. 2, the middle blue curve is an illustration of the Lorentzian response curve of the whole extended cavity. The gain bandwidth $\gamma_{a}$ can be evaluated by the convolution product of laser diode gain spectrum and Faraday atomic filter transmission spectrum, i.e., 45 MHz. Consequently, the calculated coefficient of $\Gamma_{c}/\gamma_{a}$ is 6.6, which means the active Faraday optical frequency standard works in the bad-cavity regime.

\begin{figure}[htb]
\centerline{\includegraphics[width=8cm]{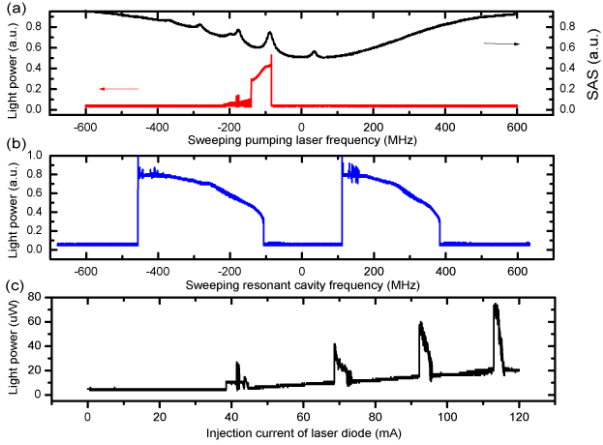}}
\caption{ (Color online) The intensity characteristics and threshold behaviors of the active Faraday optical frequency standard. (a) The variations of the emitted light power while sweeping pumping laser  frequency of the Faraday atomic filter; (b) The variations of the emitted light power  while sweeping resonant cavity frequency, i.e. the extended cavity length; (c) The variations of the emitted light power while increasing the injection current of the laser diode. }
\end{figure}

When the Faraday atomic filter is placed inside the extended cavity of laser diode, and the cavity mirror M$_{0}$ is adjusted to form feedback, the stimulated emission of the active Faraday optical frequency standard is realized. The intensity characteristics and threshold behaviors are illustrated in Fig. 3. In order to show the dependence of the stimulated emission on the Faraday atomic filter, the frequency of pump laser is unlock temporarily and swept. The emitted light power is recorded by a photodiode (PD$_{1}$) shown in Fig. 3(a). When the pumping laser is at the frequency of Cs 6 $^{2}S_{1/2}$ $F$ = 4 to 6 $^{2}P_{3/2}$ $F'$ = 4 and 5 crossover line, there is stimulated emission due to the transmission of the Faraday atomic filter. Fig. 3(b) shows the variations of the emitted light power while sweeping the resonant extended cavity frequency, calibrated by the voltage and piezoelectric coefficient of the PZT. When changing the resonant extended cavity mode frequency, the emitted light frequency is shifted away from the center of the Faraday atomic filter due to the cavity pulling effect. The stimulated emission would be extinguished once the emitted light frequency is beyond the bandwidth of the Faraday atomic filter. The frequency range of the stimulated emission working is measured to be 350(20) MHz. Fig. 3(c) shows the emitted light power varying with the injection current of the laser diode. There are four thresholds of the injection current which could be explained similarly to Fig. 3(b) considering the variation of semiconductor refractive index \cite{Zhuang11}. The maximum output power is about 75 $\upmu$W which is sufficient for phase locking.

\begin{figure}[htb]
\centerline{\includegraphics[width=8cm]{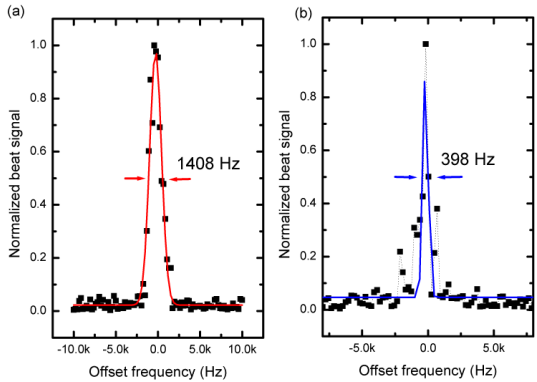}}
\caption{The optical frequency charactistics of the active Faraday optical frequency standard when beat between two seperate setups with FWHM (a)1408 Hz and (b) 398 Hz.}
\end{figure}

The optical emission frequency linewidth can be evaluated using the Schawlow-Townes formula $\Delta f_{o}=hf_{o}\Gamma_{c}^{2}/[4\pi P_{o}(1+\Gamma_{c}/\gamma_{a})^{2}]$ where $f_{o}$ is the emitted light frequency, $P_{o}$ is the emitted light power. Based on the parameters above, the quantum-limited linewidth of the active Faraday optical frequency standard is 0.33(2) Hz. In Fig. 4, we demonstrate the optical emission frequency linewidth by the heterodyne signal between two independent identical setups. The beat signal from the spectrum analyzer is normalized for fitting. The RBW is set at 100 Hz, and the swept frequency range 20 kHz within 200 ms. In Fig. 4, the data points are well fitted by a Gaussian profile (solid line), which indicates that the linewidth is dominated by 1/f noise and we can deduce the linewidth for one setup by the measured FWHM divided by $\sqrt{2}$. From Fig. 4 (a), we find the FWHM of the beat signal to be 1408(37) Hz, which means the FWHM linewidth for each setup is 996(26) Hz. This result is nearly two orders better than frequency stabilized laser with interference filter \cite{Baillard06} or atomic filter \cite{Miao11} working in good-cavity regime, which indicates that the active Faraday optical frequency standard working in bad-cavity regime can reduce the influence of environmental noises on frequency stability dramatically. An even better result with FWHM of the beat signal fitted to be 398(32) Hz and the FWHM linewidth 281(23) for each setup is shown in Fig. 4(b). The outlier data points indicate that vibrations still have great influences on frequency stability. Above all, the actual linewidth of the active Faraday optical frequency standard is coming closer to the quantum-limited linewidth.

Moreover, we heterodyned one output light beam of the active Faraday optical frequency standard with the pumping laser. The beat signal from APD$_{2}$ centers at 113MHz which means the frequency of output light equals that of the pumping laser, i.e., the frequency of the active Faraday optical frequency standard is determined by that of the Faraday atomic filter. On the other hand, the cavity pulling of the optical emission frequency is measured by changing the resonant extended cavity frequency within 350(20) MHz similarly to Fig. 3(a). The center frequency of the beat signal varies within a range of 40(5) MHz corresponding to FWHM bandwidth of the Faraday atomic filter. The ratio of 40(5) to 350(20) MHz indicates that the cavity pulling coefficient $P_{c}\equiv\delta f_{o}/\delta f_{c}=1/(1+\Gamma_{c}/\gamma_{a})$ is 0.11(2) and the coefficient of  $\Gamma_{c}/\gamma_{a}$  is 7.8(2). The results confirm that the active Faraday frequency standard posses narrower frequency linewidth and better frequency stability with help of suppressed cavity pulling effect.

It should be noted that there are not any vibration isolations in our experiment, two setups are arranged to share the same Faraday atomic filter and the same cavity mirror on purpose to get rid of large-scale noises. Therefore, the frequency stability is still limited to environmental noises and the slow frequency drift makes it hard to give the Allan deviation of our system. Moreover, the drift of the pumping laser frequency would cause frequency drift of the active Faraday optical frequency standard and even result in extinguishment of the stimulated emission. In the future, vibration isolation will be adopt and the extended cavity will be stabilized by electronic servo loops. We will also consider the possibility to realize a narrower bandwidth Faraday atomic filters without pumping lasers, like calcium, strontium, ytterbium Faraday atomic filters in optical lattices at magic wavelengthes. The gain medium can be provided by Ti: sapphire and dye, besides semiconductor materials.

In conclusion, we propose and demonstrate active Faraday frequency standards using the cesium 852 nm narrow bandwidth Faraday atomic filter in the extended cavity of the diode laser. The emitted light power reaches a maximum of 75 $\upmu$W. The optical emission frequency is stabilized to the cesium 6 $^{2}S_{1/2}$ $F$ = 4 to 6 $^{2}P_{3/2}$ $F'$ = 4 and 5 crossover transition line. The beat signal between two independent identical setups indicates a linewidth of 996(26) Hz by Gaussian fitting, which is 5.3$\times$10$^{3}$ times smaller than the natural linewidth of the cesium 852 nm transition line. The results presented here open up a new opportunity of building active Faraday optical clocks without local super-narrow linewidth lasers. With Faraday effect \cite{Faraday} at clock transition, higher performance of frequency stability can be expected for the future active Faraday optical clocks.

We thank Xiaobo Xue and Mo Chen for language assistance. This work is supported by National Natural Science Foundation of China under No. 10874009 and 11074011, and International Science $\&$ Technology Cooperation Program of China under No. 2010DFR10900.

\pagebreak


\begin{thebibliography}{99}


\bibitem{Takamoto05} M. Takamoto, F. -L. Hong, R. Higashi, and H. Katori, ``An optical lattice clock,'' Nature, \textbf{435,} 321 (2005).
\bibitem{Hinkley13}N. Hinkley, J. A. Sherman, N. B. Phillips, M. Schioppo, N. D. Lemke, K. Beloy, M. Pizzocaro, C. W. Oates, and
A. D. Ludlow, ``An atomic clock with 10$^{-18}$ instability,'' Science, \textbf{341}, 1215 (2013).
\bibitem{Ye14} B. J. Bloom, T. L. Nicholson, J. R. Williams, S. L. Campbell, M. Bishof, X. Zhang, W. Zhang, S. L. Bromley and J. Ye, ``An optical lattice clock with accuracy and stability at
the 10$^{-18}$ level,'' Nature \textbf{506,} 71 (2014).
\bibitem{Katori} I. Ushijima, M. Takamoto, M. Das, T. Ohkubo, H. Katori, ``Cryogenic optical lattice clocks with a relative frequency difference of $1\times 10^{-18}$,'' arXiv:1405.4071.

\bibitem{Jiang11} Y. Y. Jiang, A. D. Ludlow, N. D. Lemke, R. W. Fox, J. A. Sherman, L.-S. Ma and C. W. Oates, ``Making optical atomic clocks more stable with
10$^{-16}$-level laser stabilization,'' Nature Photon., \textbf{5,} 158 (2011).
\bibitem{Kessler12} T. Kessler, C. Hagemann, C. Grebing, T. Legero, U. Sterr, F. Riehle, M. J. Martin, L. Chen and J. Ye, ``A sub-40-mHz-linewidth laser based on a silicon single-crystal
optical cavity,'' Nature Photon., \textbf{6,} 687 (2012).
\bibitem{Chen05} J. -B. Chen and X. -Z. Chen, ``Optical Lattice Laser,'' Proceedings of the 2005 IEEE International Frequency Control Symposium and Exposition (IEEE, New York, 2005), p. 608.
\bibitem{Chen09} J. -B. Chen, ``Active optical clock,'' Chin. Sci. Bull., \textbf{54,} 348 (2009).
\bibitem{Chen08} D. -S. Yu and J. -B. Chen, ``Laser theory with finite atom-field interacting time,'' Phys. Rev. A., \textbf{78,} 013846 (2008).
\bibitem{Chen11} W. Zhuang and J. -B. Chen, ``Feasibility of extreme ultraviolet active optical
clock,'' Chin. Phys. Lett., \textbf{28,} 080601 (2011).
\bibitem{Kazakov13} G. A. Kazakov and T. Schumm, ``Active optical frequency standard using sequential coupling of atomic ensembles,'' Phys. Rev. A., \textbf{87,} 013821 (2013).
\bibitem{Meiser09} D. Meiser, J. Ye, D. R. Carlson, and M. J. Holland, ``Prospects for a millihertz-linewidth
laser,'' Phys. Rev. Lett., \textbf{102,} 163601 (2009).
\bibitem{Meiser10} D. Meiser and M. J. Holland, ``Steady-state superradiance with alkaline-earthmetal
atoms,'' Phys. Rev. A., \textbf{81,} 033847 (2010).
\bibitem{Bohnet12} J. G. Bohnet, Z. -L. Chen, J. M. Weiner, D. Meiser, M. J. Holland, and J. K. Thompson, ``A steady-state superradiant laser with less than one
intracavity photon,'' Nature, \textbf{484,} 78 (2012).
\bibitem{Sterr09} U. Sterr and C. Lisdat, ``Millihertz-linewidth lasers: A sharper laser,'' Nature Phys., \textbf{5,} 382 (2009).
\bibitem{Kuppens94} S. J. M. Kuppens, M. P. Van Exter, and J. P. Woerdman, ``Quantum-limited linewidth of
a bad-cavity laser,'' Phys. Rev. Lett., \textbf{72}, 3815 (1994).
\bibitem{Kuppens96} S. J. M. Kuppens, M. P. Van Exter, J. P. Woerdman, and M. I. Kolobov, ``Observation of the effect of
spectrally inhomogeneous gain on the quantum-limited laser linewidth,'' Opt. Commun., \textbf{126,} 79 (1996).
\bibitem{Faraday} M. Faraday,"Experimental Researches in Electricity. Nineteenth Series," Phil. Trans. R. Soc. Lond., \textbf{136,} 1 (1846).
\bibitem{Turner02} L. D. Turner, V. Karaganov, and P. J. O. Teubner, ``Sub-Doppler bandwidth atomic optical filter,'' Opt. Lett., \textbf{27,} 500 (2002).
\bibitem{Cer09} A. Cer\`{e}, V. Parigi, M. Abad, F. Wolfgramm, A. Predojevic, and M. W. Mitchell, ``Narrowband tunable filter based on
velocity-selective optical pumping
in an atomic vapor,'' Opt. Lett., \textbf{34,} 1012 (2009).
\bibitem{Popescu10} A. Popescu and T. Walther, ``On an ESFADOF edge-filter for a range resolved Brillouin lidar: The high vapor density and high pump intensity regime,'' Appl. Phys. B, \textbf{98,} 667 (2010).
\bibitem{Liu12} S. -Q. Liu, Y. -D. Zhang, H. Wu, and P. Yuan, ``Ultra-narrow bandwidth atomic filter based on optical-pumping-induced dichroism
realized by selectively saturated absorption,'' Opt. Commun., \textbf{285,} 1181 (2012).
\bibitem{Wang12} Y. -F. Wang, S. -N. Zhang, D. -Y. Wang, Z. -M. Tao, Y. -L. Hong, and J. -B. Chen, ``Nonlinear optical filter with ultranarrow bandwidth
approaching the natural linewidth,'' Opt. Lett., \textbf{37,} 4059 (2012).
\bibitem{Chen87} X. Chen, V. Telegdi, and A. Weis, "Magneto-optical rotation near the caesium D2 line(Macaluso-Corbino effect) in intermediate fields: I. Linear regime," J. Phys. B: At. Mol. Phys., \textbf{20,} 5653 (1987).
\bibitem{Sorokin69} P. P. Sorokin, J. R. Lankard, V. L. Moruzzi, and A. Lurio, ``Frequency-locking of organic dye lasers to atomic resonance lines,'' Appl. Phys. Lett., \textbf{15,} 179 (1969).
\bibitem{Zhang13} X. -G. Zhang, Z. M. Tao, C. -W. Zhu, Y. -L. Hong, W. Zhuang and J. -B. Chen, ``An all-optical locking of a semiconductor
laser to the atomic resonance line with 1
MHz accuracy,'' Opt. Express, \textbf{21,} 28010 (2013).
\bibitem{Miao11} X. Miao, L. Yin, W. Zhuang, B. Luo, A. Dang, J. Chen, and H. Guo, ``Demonstration of an external-cavity diode
laser system immune to current and temperature fluctuations,'' Rev. Sci. Instrum., \textbf{82,} 086106 (2011).
\bibitem{Zhuang11} W. Zhuang, D. -S. Yu, Z. -W. Liu, and J. -B. Chen, ``Multi-threshold second-order phase transition,'' Chin. Sci. Bull., \textbf{56,} 3812(2011).
\bibitem{Baillard06} X. Baillard, A. Gauguet, S. Bize, P. Lemonde, P. Laurent, A. Clairon, and P. Rosenbusch, ``Interference-filter-stabilized external-cavity diode lasers,'' Opt. Commun., \textbf{266,} 609 (2006).

\end{thebibliography}
\end{document}